\documentclass[reprint,amsmath,amssymb]{revtex4-2}

\usepackage{graphicx}
\usepackage[utf8]{inputenc}
\usepackage[T1]{fontenc}
\usepackage{bm}
\usepackage{booktabs}

\begin{document}

\title{What causes the magnetic curvature drift?}

\author{Johnathan K. Burchill}
\email{j.burchill@ucalgary.ca}
\affiliation{Department of Physics and Astronomy, University of Calgary, Calgary, Canada}

\date{\today}

\begin{abstract} 

    When asked what causes the magnetic curvature drift of a charged-particle
    moving in a curving magnetic field, people respond that there is an
    `F-cross-B' motion of the `guiding center' due to the centrifugal force on
    the particle as it follows the magnetic field line. This and similar
    explanations `beg the question' by assuming that the particle follows the
    field line. In a curving magnetic field, however, a particle moving
    parallel to the field direction soon won't be. The convective rotation of
    the field along the particle trajectory ensures that the Lorentz force
    switches on, and the resulting acceleration rotates the velocity vector
    back into alignment periodically. The gyration is not symmetric about the
    field vector, and the resulting velocity offset is the curvature drift.
    This explanation is guided by Newton's second law of motion in vector
    notation. It provides a common framework for explaining the three
    guiding-center motions of a charged particle in a static nonuniform
    magnetic field: curvature drift, mirror reflection in a magnetic bottle,
    and gradient-B drift. The discussion aims to provide
    insight to instructors of electricity and magnetism or plasma physics at
    the intermediate- to advanced-undergraduate level. 

\end{abstract}

\maketitle

\section{Introduction}
\label{sec:intro}

Kruskal (1965) gives a clear introduction to charged-particle motion in a
magnetic field with a focus on the guiding-center
approximation\cite{Kruskal1965}. There, as in introductory textbooks of plasma
physics, the magnetic curvature drift is derived from the centrifugal force
$F_c = m v^2_b \kappa$ experienced by a particle moving with speed $v_b =
\mathbf{v}\cdot \hat{b}$ along a field line with curvature $\boldsymbol{\kappa}
=\kappa\hat{\kappa}$. Under the guiding-center approximation for which the
gyromotion is averaged out, the particle
`F-cross-B' drifts as
\begin{equation} 
    \langle \mathbf{v}_\perp\rangle \sim \frac{\mathbf{F}\times\mathbf{B}}{q B^2}, 
    \label{eq:f-cross-b} 
\end{equation}
where $q$ is the particle's electric charge, $\mathbf{B}=B\hat{b}$ is the
magnetic field vector, and $\mathbf{F}$ is any force not associated with the
zeroth-order Lorentz force gyromotion. In the case of the centrifugal force this yields
\begin{equation} 
    \mathbf{v}_{\perp,R} = \frac{m v^2_b}{q B^2}\mathbf{B}\times\boldsymbol{\kappa}. 
    \label{eq:classic-curvature-drift} 
\end{equation}

The mirror effect in a magnetic bottle (like a dipole field) is explained as a
consequence of the constancy of the magnetic moment
(the first of the adiabatic invariants of gyromotion):
\begin{equation}
    \mu = \frac{m v^2_\perp}{2 B},
    \label{eq:magnetic-moment}
\end{equation} 
where $v_\perp=|\mathbf{v}-v_b\hat{b}|$. 
The gyrating charge has a dipole moment $\boldsymbol{\mu}=-\mu \hat{b}$, 
so the particle is pushed on average in the direction of weaker magnetic field:
\begin{equation}
    \langle F_b \rangle = -\mu \nabla_b B,
    \label{eq:classic-mirror-force}
\end{equation}
where $\nabla_b$ denotes the parallel component of the gradient.
The magnetic Lorentz force is perpendicular to $\mathbf{v}$ and
does no work, so $|\mathbf{v}|^2 = v_b^2 + v_\perp^2 = v^2$ is a constant of
the motion. 
For pitch angle $\alpha = \cos^{-1}(v_b/v)$, the ratio $\sin^2\!\alpha / B$ is
approximately constant along the trajectory. 

The gradient-B drift 
\begin{equation} 
\mathbf{v}_{\perp,\nabla B} = \frac{m v^2_\perp}{2q B^3}\mathbf{B}\times\nabla B
    \label{eq:class-grad-b-drift}
\end{equation} 
can be explained by noting that the particle's gyroradius $r_L = |m v_\perp/qB|$
varies inversely as the field magnitude along the orbit. A notable alternative
invokes energy conservation and the first adiabatic invariant\cite{Cully1999}. 

The guiding-center approximation is a split: fast gyration about a slowly
drifting point, the center of gyration. This works as long as
the Larmor radius is small compared to the field gradient 
scale, as expressed by a `smallness parameter'~\cite{CaryBrizard2009}
\begin{equation}
\varepsilon = \left|r_L \nabla_\perp B/B\right| \ll 1.
    \label{eq:smallness-parameter}
\end{equation}
At zeroth order the gradient vanishes and the motion is purely
helical. First order in $\varepsilon$ captures linear trends in the field
gradient around the orbit.
It is in this regime that Eqs.~(\ref{eq:f-cross-b})
through (\ref{eq:class-grad-b-drift}), together with gyration, are valid
descriptions of the motion\cite{Kruskal1965} when the magnetic field and 
its gradient are evaluated at the position $\mathbf{R}$ of the guiding center. 

\section{A pedagogical puzzle}
\label{sec:the-problem}
The centrifugal force explanation of curvature drift raises an interesting
puzzle as illustrated with the following dialogue: 

INSTRUCTOR: What causes the magnetic curvature drift?

STUDENT: The centrifugal force experienced by the particle as it moves along a
curved field line.

INSTRUCTOR: What real force pushes the particle?

STUDENT: The centripetal force. 

INSTRUCTOR: Can you elaborate? What fundamental force accelerates the particle? 

STUDENT: It must be the Lorentz force. The drift is required to ensure that the 
particle follows the field line.

INSTRUCTOR: Okay. Suppose the particle happens to be moving exactly parallel to
the magnetic field vector, so the angle between the velocity and the magnetic
field is zero. The Lorentz force is zero. What causes the particle to
curve along the magnetic field line?

At this point students are baffled: if the first adiabatic invariant
(Eq.~\ref{eq:magnetic-moment}) is zero, presumably it remains so, and the guiding
center follows the field line. How is this possible if the Lorentz force is
zero? It's a puzzle that draws students in. Solving it means figuring out how
the Lorentz force varies along the particle's trajectory. The explanation, as
we will see, reaches into the mirror force and gradient-B drift effects too.

\section{It must be the Lorentz force} 
\label{sec:physics} 

The physics is contained in Newton's second law of motion: 
\begin{equation}
    \frac{d\mathbf{p}}{dt} = q \mathbf{v}\times\mathbf{B},
    \label{eq:LorentzForce} 
\end{equation}
the field evaluated at the position of the particle. 
The particle's momentum can be changed only by the Lorentz force. If the
particle is moving exactly parallel to the field direction, the Lorentz force
is zero, but it soon won't be. The field rotates along the
particle's trajectory. This turns on and varies the Lorentz force, 
and gyration ensues. As the particle moves through the curving field, the Lorentz
force rotates the particle's velocity back into alignment with the field.
The rotation of the velocity vector is periodic but not symmetric about the
field line. The offset is the curvature drift. The change in pitch
angle is usually very small and in simulations it can look like the particle
follows the field line exactly. 

Let's examine this for non-relativistic motion in a static nonuniform magnetic
field. The Lorentz force varies along the trajectory due to a combination of
changing velocity and changing magnetic field, and this drives an acceleration
of the velocity vector:
\begin{equation}
    m\ddot{\mathbf{v}} = q \dot{\mathbf{v}}\times\mathbf{B} + q \mathbf{v}\times\dot{\mathbf{B}}.
    \label{eq:newton-time-deriv}
\end{equation}

Newton's second law $\dot{\mathbf{v}} = (q/m)\mathbf{v}\times \mathbf{B}$ can
be used to eliminate the first time derivative of the velocity, and the total
time derivative of the magnetic field can be expressed in terms of a convective
derivative:
\begin{equation}
    \dot{\mathbf{B}} = \frac{\partial \mathbf{B}}{\partial t} + \mathbf{v}\cdot\nabla \mathbf{B},
    \label{eq:b-dot}
\end{equation}
where the partial time derivative is zero in the static case treated here.
With the first substitution, the term 
$q\dot{\mathbf{v}}\times\mathbf{B}$ in Eq.~(\ref{eq:newton-time-deriv}) reduces to
\begin{equation}
\begin{aligned}
    q\dot{\mathbf{v}}\times\mathbf{B} &= \frac{q^2}{m}(\mathbf{v}\times\mathbf{B})\times\mathbf{B} \\
                                      &= \frac{q^2}{m}\left[(\mathbf{v}\cdot\mathbf{B})\mathbf{B} - B^2\mathbf{v}\right] \\
                                      &= -\frac{q^2 B^2}{m}\,\mathbf{v}_\perp,
    \label{eq:double-cross-identity}
\end{aligned}
\end{equation}
the parallel piece $B^2 v_b\hat{b}$ canceling between the two terms in the
bracket. The $q\mathbf{v}\times\dot{\mathbf{B}}$ term in
Eq.~(\ref{eq:newton-time-deriv}) carries the field-gradient information through
the field-magnitude/field-direction split
\begin{equation}
    \mathbf{v}\cdot\nabla\mathbf{B} = (\mathbf{v}\cdot\nabla B)\,\hat{b} + B\,(\mathbf{v}\cdot\nabla\hat{b}),
    \label{eq:B-grad-split}
\end{equation}
which is the organizing decomposition for everything that follows.
This allows us to express Eq.~(\ref{eq:newton-time-deriv}) as
\begin{equation}
    \ddot{\mathbf{v}}_\perp = - \Omega^2 \mathbf{v}_\perp + \frac{\Omega}{B}\mathbf{v}\times\hat{b}(\mathbf{v}\cdot\nabla B) + \Omega\left[ \mathbf{v}\times(\mathbf{v}\cdot\nabla\hat{b}) \right]_\perp
    \label{eq:eom-perp}
\end{equation}
with angular frequency $\Omega = q B/m$ whose sign is determined by $q$,
and, using $\dot{p}_b = \frac{d}{dt}(\mathbf{p}\cdot\hat{b}) = \dot{\mathbf{p}}\cdot\hat{b} + \mathbf{p}\cdot\dot{\hat{b}}$ with $\dot{\mathbf{p}}\cdot\hat{b} = 0$ and $\dot{\hat{b}} = \mathbf{v}\cdot\nabla\hat{b}$,
\begin{equation}
    \dot{p}_b = m\mathbf{v}\cdot (\mathbf{v}\cdot\nabla\hat{b}).
    \label{eq:N2-parallel}
\end{equation}

The angular frequency in Eq.~(\ref{eq:eom-perp}) depends on the magnetic field strength, which varies around the orbit. Eq.~(\ref{eq:eom-perp}) can be cast as a driven harmonic oscillator using $B = B_0 + \delta B$ where $B_0$ is the magnetic field intensity at the center of gyration:
\begin{equation}
    \ddot{\mathbf{v}}_\perp = - \Omega^2_0 \left[ \mathbf{v}_\perp - \mathbf{v}_{\nabla B}(\mathbf{v}) - \mathbf{v}_{\nabla \hat{b}}(\mathbf{v}) \right].
    \label{eq:driven-oscillator}
\end{equation}
The homogeneous solution is gyration about
the field vector at angular frequency $\Omega_0 = q B_0 / m$. 
The inhomogeneous driving has been split into two parts. The first comes 
about from the convective rate of change of the field magnitude
\begin{equation}
    \mathbf{v}_{\nabla B}(\mathbf{v}) = \frac{m}{q B^2_0} \mathbf{v}\times\hat{b} (\mathbf{v} \cdot \nabla B) - \left(2\frac{\delta B}{B_0} + \frac{\delta B^2}{B^2_0}\right)\mathbf{v}_\perp,
    \label{eq:field-magnitude-velocity-driver}
\end{equation}
while the second arises from the convective rate of change of the field direction:
\begin{equation}
    \mathbf{v}_{\nabla \hat{b}}(\mathbf{v}) = \frac{m B}{q B^2_0} \left[\mathbf{v} \times (\mathbf{v} \cdot \nabla\hat{b})\right]_\perp
    \label{eq:field-direction-velocity-driver}
\end{equation}
with $[\mathbf{X}]_\perp \equiv \mathbf{X} - (\mathbf{X}\cdot\hat{b})\hat{b}$ the perpendicular projection.
Eq.~(\ref{eq:driven-oscillator}) governs the perpendicular part of $\mathbf{v}$; Eq.~(\ref{eq:N2-parallel}) the parallel part.

Eqs.~(\ref{eq:N2-parallel}), (\ref{eq:field-magnitude-velocity-driver}) and
(\ref{eq:field-direction-velocity-driver}) are exact expressions whose orbit
averages are the magnetic mirror force, the gradient-B drift, and the curvature
drift. 

\section{Field-direction convective derivative}
\label{sec:drift-and-mirror}

It is instructive to investigate the structure of the field-direction convective derivative vector
\begin{equation}
    \mathbf{v}\cdot\nabla\hat{b}. 
    \label{eq:convective-field-derivative}
\end{equation}
Let's work in a coordinate system that is defined by the field direction
and its curvature:
\begin{align}
    \hat{\kappa} &= \hat{b}\cdot\nabla\hat{b}/\kappa \notag \\
    \hat{\tau} &= \hat{b}\times\hat{\kappa} \label{eq:frenet-serret-coords} \\
    \hat{b} &= \mathbf{B}/B. \notag
\end{align}
This coordinate system is sometimes referred to as the Frenet-Serret frame,
defined in terms of the principal normal $\hat{\kappa}$ (along
the bending direction, with magnitude $\kappa$), the binormal $\hat{\tau}$,
and the tangent $\hat{b}$ of the magnetic
field line.
The concept of a field line is superfluous here: only the field direction 
and its derivatives at a point are needed. The particle's velocity in this frame is
\begin{equation}
    \mathbf{v} = v_\kappa \hat{\kappa} + v_\tau \hat{\tau} + v_b \hat{b}.
    \label{eq:particle-velocity}
\end{equation}
To calculate Eq.~(\ref{eq:convective-field-derivative}) we need the 
components of the field-direction gradient tensor $\nabla \hat{b}$:
\begin{align}
    \hat{\kappa}\cdot\nabla\hat{b} &= G_{\kappa \kappa} \hat{\kappa} + G_{\tau \kappa} \hat{\tau} \notag \\
    \hat{\tau}\cdot\nabla\hat{b} &= G_{\kappa \tau} \hat{\kappa} + G_{\tau \tau} \hat{\tau} \label{eq:field-direction-gradient-vectors} \\
    \hat{b}\cdot\nabla\hat{b} &= \kappa \hat{\kappa} \notag 
\end{align}
where the field-direction gradient tensor elements
    \begin{equation}
    G_{mn} = \hat{m}\cdot(\nabla\hat{b}\cdot\hat{n})
    \label{eq:Gij_elements}
\end{equation}
represent the $m$-component of the rate of change of the field direction in the
$\hat{n}$ direction. There are only five non-zero terms in
Eq.~(\ref{eq:field-direction-gradient-vectors}) because the length of the field
direction unit vector is constant. The constraint $|\hat{b}|^2 =
1$ forces $\hat{b}\cdot\nabla\hat{b}$ to be perpendicular to $\hat{b}$ in every
direction, eliminating the three components $G_{b\kappa}$, $G_{b\tau}$, and
$G_{bb}$; the choice of $\hat{\kappa}$ along $\hat{b}\cdot\nabla\hat{b}$
eliminates $G_{\tau b}$. The curvature $G_{\kappa b}$ is denoted `$\kappa$' as
before. This is the mathematical foundation of the main idea: If the particle
is moving parallel to the field direction, it soon won't be, and the Lorentz
force is going to turn on.

\section{Curvature drift}
\label{sec:curvature-drift}

Eqs.~(\ref{eq:N2-parallel}) and (\ref{eq:driven-oscillator}) exactly describe
non-relativistic charged-particle motion in a static, non-uniform magnetic
field. This motion is not always dominated by gyration. For a particle 
momentarily moving
parallel to the field direction ($\mathbf{v} = v_b \hat{b}$) in a curving
field, gyromotion ensues with a small but non-negligible $\mathbf{v}_\perp$
in Eq.~(\ref{eq:driven-oscillator}). But the particle
velocity does not rotate symmetrically about $\mathbf{B}$. The driving term
Eq.~(\ref{eq:field-direction-velocity-driver}) gives an offset:
\begin{equation}
    \left. \mathbf{v}_{\nabla \hat{b}}(\mathbf{v})\right|_{\sin\!\alpha\sim 0} \sim \frac{m}{q B_0} v_b \hat{b} \times (v_b \hat{b}\cdot \nabla\hat{b}),
    \label{eq:curvature-drift-driver-small-angle}
\end{equation}
with $B\simeq B_0$. Both copies of $\mathbf{v}$ in
Eq.~(\ref{eq:field-direction-velocity-driver}) have been replaced by
$v_b\hat{b}$. Using $\hat{b}\cdot\nabla\hat{b} = \kappa\hat{\kappa}$ 
(Eq.~\ref{eq:field-direction-gradient-vectors}) and
$\hat{b}\times\hat{\kappa} = \hat{\tau}$
(Eq.~\ref{eq:frenet-serret-coords}), this becomes
\begin{equation}
    \left. \mathbf{v}_{\nabla \hat{b}}(\mathbf{v})\right|_{\sin\!\alpha\sim 0} \sim \frac{m v_b^2 \kappa}{q B_0}\,\hat{\tau},
    \label{eq:curvature-drift-tau}
\end{equation}
already perpendicular to $\hat{b}$, so the projection $[\,\cdot\,]_\perp$ in
Eq.~(\ref{eq:field-direction-velocity-driver}) is trivial here. The particle
does not follow the field line, not even on average. This is magnetic curvature
drift. Rewriting in terms of $\mathbf{B} = B\hat{b}$ and the
curvature vector $\boldsymbol{\kappa} = \kappa\hat{\kappa}$ recovers
the classic form
\begin{equation}
    \left. \mathbf{v}_{\nabla \hat{b}}(\mathbf{v})\right|_{\sin\!\alpha\sim 0} \sim \frac{m v^2_b}{q B^2} \mathbf{B}\times \boldsymbol{\kappa}.
\end{equation}

STUDENT: The textbook derivation of the curvature drift is much shorter. 
Why are we doing all this instead?

INSTRUCTOR: To show that we can explain the drift without assuming the answer. What 
does ``the particle moves along the field line'' mean?

STUDENT: Its velocity is parallel to $\hat{b}$.

INSTRUCTOR: For the real particle, or for the guiding center?

STUDENT: The guiding center.

INSTRUCTOR: The particle's velocity is not parallel to $\mathbf{B}$ except
instantaneously. A frame moving along the field line is moving with the guiding
center, whose path should be established by the explanation. The centrifugal
argument helps itself to the guiding-center picture and computes a correction
to it. While rigorously correct as a derivation, the explanation itself is
incomplete.

STUDENT: And the guiding-center concept itself assumes the motion is gyration 
about a slowly moving center.

INSTRUCTOR: It does. We call this an \emph{ansatz}: we conjecture the form of 
the answer and check it for consistency, order by order 
in the smallness parameter \cite{CaryBrizard2009}. 

 STUDENT: And the framework here? 

INSTRUCTOR: We are doing something similar by expressing the equation of motion as
a driven oscillator whose homogeneous solution is helical motion. The
insight is that the source of the curvature drift can be expressed directly 
in the equation of motion through the field-direction convective derivative. 
To bring this into focus, let's 
consider a cylindrically-symmetric curved magnetic field:
\begin{equation} 
    \bm{B} = B_0(-y\,\hat{x}+x\,\hat{y}+0\,\hat{z})/\sqrt{x^2+y^2}.
    \label{eq:curving-field-model}
\end{equation} 
This model isolates the curvature drift from other effects like the mirror force
and the gradient-B drift. 

\begin{figure*} 
\includegraphics[width=\textwidth]{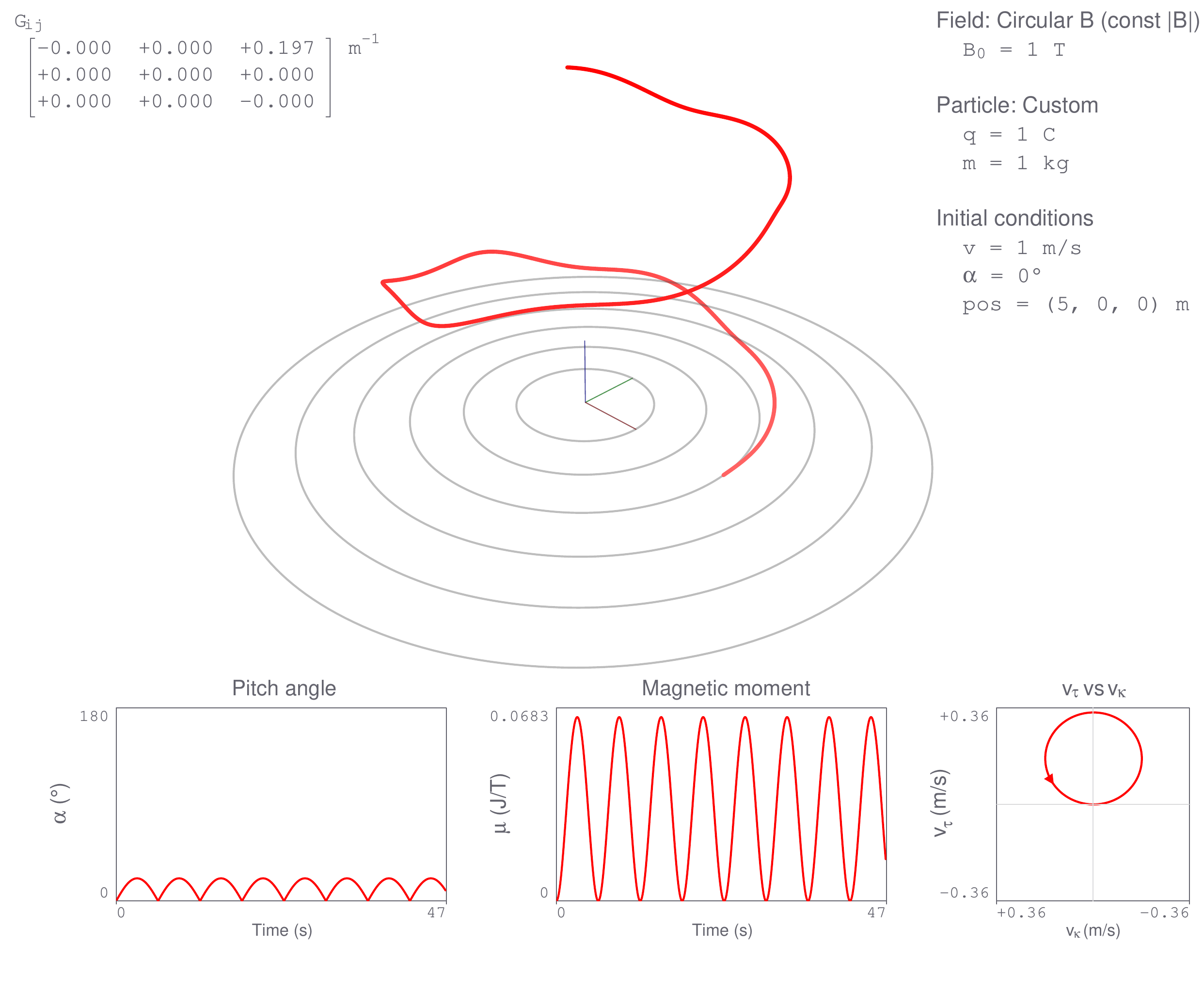} 
\caption{Lorentz Tracer\cite{LorentzTracer}
    illustration of the pure curvature
    drift of a test charge in the magnetic field of
    Eq.~(\ref{eq:curving-field-model}). Since $|\bm{B}|=B_0$ is constant, there
    is no grad-$B$ drift and no mirror force; the vertical drift is pure
    curvature drift. Bottom panels show time series of the pitch angle $\alpha$
    (left) and first adiabatic invariant $\mu$
(center). The
velocity space trajectory in a plane perpendicular to $\mathbf{B}$ is shown at
bottom right. The offset of the loop in velocity space is the curvature drift.} 
\label{fig:lorentz-tracer}
\end{figure*} 

Figure~\ref{fig:lorentz-tracer} is a screenshot from Lorentz
Tracer\cite{LorentzTracer} depicting the trajectory of a test charge in the
field given by Eq.~(\ref{eq:curving-field-model}). The initial conditions and
field parameters have been chosen to exaggerate the curvature drift: the
particle rises as the velocity and field rotate into alignment periodically.
The particle starts out moving parallel to $\mathbf{B}$ 
($\alpha = 0$, moving away from the observer), but the
pitch angle wobbles periodically between zero and a maximum value. The first
adiabatic invariant varies periodically. Gyration is evident in the
$v_\tau$--$v_\kappa$ phase diagram at lower right, centered on the curvature
drift rather than the magnetic field direction. The particle's path resembles
a helix. (A dramatic lecture illustration is to show something like
Figure~\ref{fig:lorentz-tracer} without the field lines, explain that the red
curve represents the motion of a positively-charged particle in a magnetic
field, ask `Which way does the magnetic field point?' and reveal the field
lines after students have an opportunity to voice their opinions.)

Field curvature $\kappa$ introduces variations in the 
particle's momentum along $\hat{b}$ as well. According to Eq.~(\ref{eq:N2-parallel}):
\begin{equation}
    \dot{v}_b = v^2_\kappa G_{\kappa \kappa} + v^2_\tau G_{\tau \tau} + v_\kappa v_\tau (G_{\kappa \tau} + G_{\tau \kappa}) + v_b v_\kappa \kappa.
\label{eq:large-angle-parallel-moment-change}
\end{equation}
Variations in $v_\kappa$ perturb the parallel velocity at the gyrofrequency. A
predominantly $2\Omega_0$ `flutter' in the parallel velocity arises from
non-gyrotropic field-direction variations in the $v_\kappa v_\tau
(G_{\kappa\tau} + G_{\tau\kappa})$ term. Its amplitude is greatest for
pitch angles near $\pi/4$ and $3\pi/4$. Geometrically,
$G_{\kappa\tau} + G_{\tau\kappa}$ measures the torsion-like twisting of the
curvature vector about the field direction. The curvature drift itself diminishes
as $\cos^2\alpha$ and becomes negligible when $\alpha \sim \pi/2$. The exact
form Eq.~(\ref{eq:field-direction-velocity-driver}) drives oscillations in the
$\kappa-\tau$ plane that modify the pitch angle periodically. 

\section{Mirror effect}
\label{sec:mirror-effect}

Because $v^2_\kappa$ and $v^2_\tau$ are generally not zero, the
$G_{\kappa \kappa}$ and $G_{\tau \tau}$ terms in Eq.~(\ref{eq:large-angle-parallel-moment-change}) 
systematically change the projection of the momentum along the field direction.
This is the origin of the mirroring of a particle in a magnetic
bottle, which is a kinematic effect due to field rotation: it is not caused by 
an actual force. The instantaneous Lorentz force varies as a result. Its orbit average does, however, have 
a component along the direction of the \textit{orbit-averaged} magnetic field vector. 
The mean field direction is often what is meant by the `parallel' direction.
At leading order the perpendicular motion is gyrotropic, 
    $v_\kappa = v_\perp\cos(\Omega_0 t + \phi_0)$ and 
    $v_\tau = -v_\perp\sin(\Omega_0 t + \phi_0)$, so
\begin{equation}
    \langle v_\kappa^2\rangle = \langle v_\tau^2\rangle = \tfrac{1}{2}v_\perp^2,
    \quad \langle v_\kappa v_\tau\rangle = 0,
    \label{eq:gyrotropic-averages}
\end{equation}
since $\langle\cos^2\rangle = \langle\sin^2\rangle = \tfrac{1}{2}$ and
$\langle\sin(2\Omega_0 t + 2\phi_0)\rangle = 0$ over a gyroperiod. Average
Eq.~(\ref{eq:large-angle-parallel-moment-change}) over one gyroperiod:
the gyrotropic identities above kill the cross term,
leaving
\begin{equation}
    \langle \dot{p}_b\rangle \sim m \langle v^2_\kappa G_{\kappa \kappa}\rangle + m \langle v^2_\tau G_{\tau \tau} \rangle.
    \label{eq:v-dot-para-avg}
\end{equation}
In general $G_{\kappa \kappa}$ and $G_{\tau \tau}$ are not equal. It is only by averaging over an orbit
that a connection can be made between the average parallel push and the first adiabatic
invariant using 
\begin{equation}
\begin{aligned}
    \nabla\cdot (B\hat{b}) &= \nabla B \cdot \hat{b} + B \nabla \cdot \hat{b} \\
                         &= \nabla_b B + B (G_{\kappa \kappa} + G_{\tau \tau}) \\
                         &= 0,
                         \label{eq:divb-Gij}
\end{aligned}
\end{equation}
where $\nabla\cdot\hat{b} = \mathrm{tr}(\nabla\hat{b}) = G_{\kappa\kappa} + G_{\tau\tau}$, the $G_{bb}$ 
entry vanishing by the same $|\hat{b}|^2 = 1$ constraint that reduced Eq.~(\ref{eq:field-direction-gradient-vectors}) to five terms,
and $\langle v^2_\kappa \rangle=\langle v^2_\tau\rangle=v^2_\perp/2$,
which gives
\begin{equation}
    \langle \dot{p}_b \rangle \sim -\mu \nabla_b B. \label{eq:mirror-force-classic}
\end{equation}
Because $\nabla\cdot \mathbf{B} = 0$, this `mirror force' can be expressed in terms of a
parallel gradient of the magnetic field magnitude. The physics, however, depends only on the
variation of the magnetic field direction around the orbit.

The status of $\mu$ as approximately constant in
Eq.~(\ref{eq:mirror-force-classic}) can be made explicit by
examining $\dot{\mu}$ directly. Differentiate $\mu B = \tfrac{1}{2} m v_\perp^2$
two ways: first directly, with $|\mathbf{v}|^2 = v_b^2 + v_\perp^2 = \mathrm{const}$
so that $d v_\perp^2/dt = -2 v_b \dot{v}_b$, and using Eq.~(\ref{eq:N2-parallel}) for $\dot{v}_b$,
\begin{equation}
    \frac{d(\mu B)}{dt} = -m v_b \dot{v}_b = -m v_b \mathbf{v}\cdot(\mathbf{v}\cdot\nabla\hat{b}).
    \label{eq:muB-direct}
\end{equation}
Second, by the product rule with $\dot{B} = \mathbf{v}\cdot\nabla B$ in a static field,
\begin{equation}
    \frac{d(\mu B)}{dt} = \dot{\mu} B + \mu (\mathbf{v}\cdot\nabla B).
    \label{eq:muB-product}
\end{equation}
Equating Eqs.~(\ref{eq:muB-direct}) and (\ref{eq:muB-product}) and solving for $\dot{\mu}$,
\begin{equation}
    \dot{\mu} = -\frac{\mu}{B}(\mathbf{v}\cdot\nabla B) - \frac{m v_b}{B}\mathbf{v}\cdot(\mathbf{v}\cdot\nabla\hat{b}),
    \label{eq:mu-dot-instantaneous}
\end{equation}
equivalent to Eq.~(5.30) of Cary and Brizard~\cite{CaryBrizard2009} (with $\mathbf{E} = 0$).

Both gradients of the field appear explicitly in
Eq.~(\ref{eq:mu-dot-instantaneous}), mirroring the
field-magnitude/field-direction split of Eq.~(\ref{eq:B-grad-split}). The first
term carries the convective change of the field magnitude. The second carries
the field-direction rotation: $\nabla\hat{b}$ changes the pitch angle,
transferring energy between $v_b^2$ and $v_\perp^2$.

Orbit-averaging Eq.~(\ref{eq:mu-dot-instantaneous}) and
    factoring out slowly varying quantities (leading order in $\varepsilon$, so
    that $B$, $\mu$, and the gradients $\nabla B$ and $\nabla\hat{b}$ may be
    held constant on the gyroperiod and pulled out of the average),
\begin{equation}
    \langle\dot{\mu}\rangle \approx -\frac{\mu}{B}\langle\mathbf{v}\cdot\nabla B\rangle 
    - \frac{m v_b}{B}\langle\mathbf{v}\cdot(\mathbf{v}\cdot\nabla\hat{b})\rangle.
    \label{eq:mu-dot-avg-exact}
\end{equation}
With $\langle\mathbf{v}\cdot(\mathbf{v}\cdot\nabla\hat{b})\rangle =
\tfrac{1}{2} v_\perp^2 (G_{\kappa\kappa} + G_{\tau\tau})$ (gyroaverage of
Eq.~\ref{eq:large-angle-parallel-moment-change}),
$\langle\mathbf{v}\cdot\nabla B\rangle = v_b\nabla_b B$ (terms with single factors of 
$v_\kappa$ and $v_\tau$ vanish), and $\tfrac{1}{2} m v_\perp^2 = \mu B$,
\begin{equation}
    \langle\dot{\mu}\rangle \approx -\frac{\mu v_b}{B} \big[ \nabla_b B + B (G_{\kappa\kappa} + G_{\tau\tau}) \big].
    \label{eq:mu-dot-bracket}
\end{equation}
The bracket has the form of $\nabla\cdot\mathbf{B}$ given by
Eq.~(\ref{eq:divb-Gij}), with the gradients evaluated at the gyrocenter under
the same leading-order, constant-gradient assumption that produced
Eq.~(\ref{eq:mirror-force-classic}). $\nabla\cdot\mathbf{B} = 0$ enters at this
order:
\begin{equation}
    \langle\dot{\mu}\rangle \approx -\frac{\mu v_b}{B} \nabla\cdot\mathbf{B} \approx 0,
    \label{eq:mu-dot-averaged}
\end{equation}
recovering the gyroaveraged $\langle\dot\mu_0\rangle \equiv 0$ of Cary and
Brizard~\cite{CaryBrizard2009}. The orbit average is not a universal invariant:
in fields where the gradient varies appreciably over a Larmor radius (e.g.,
dipole orbits sampling a wide range of $|\mathbf{B}|$ within a few gyroperiods)
the slow-variable approximation fails, and $\langle\dot{\mu}\rangle$ deviates
from zero even though $\nabla\cdot\mathbf{B} = 0$ everywhere.

The wobble of $\mu$ visible in Figure~\ref{fig:lorentz-tracer}
is consistent with Eq.~(\ref{eq:mu-dot-instantaneous}): $|\mathbf{B}|$ is
constant in the field of Eq.~(\ref{eq:curving-field-model}), so $\nabla B = 0$
and $\dot{\mu} = -(m v_b/B)\mathbf{v}\cdot(\mathbf{v}\cdot\nabla\hat{b})$. The
field-direction gradient does the work of changing $\mu$ here, even though
$\nabla B = 0$. The orbit average closes at leading order because $\nabla_b B =
0$ and $\nabla\cdot\mathbf{B} = 0$ together force $G_{\kappa\kappa} +
G_{\tau\tau} = 0$, so
$\langle\mathbf{v}\cdot(\mathbf{v}\cdot\nabla\hat{b})\rangle \approx 0$ and
$\langle\dot{\mu}\rangle \approx 0$. Instantaneously $\mu$ wobbles; on average
it is approximately conserved.

That the gradient of the magnetic field strength does not cause the mirror
force can be seen for an unphysical field like 
\begin{equation}
    \mathbf{B}(\mathbf{r}) = B_0 (1 + \lambda z) \hat{z}
    \label{eq:unphysical-field}
\end{equation}
where the field strength varies in the direction of the field, but
for which the field-direction gradient tensor is identically zero.
 This field has divergence $\nabla \cdot \mathbf{B} = B_0 \lambda \neq 0$.
Eq.~(\ref{eq:mirror-force-classic}) predicts a mirror effect
\begin{equation}
\begin{aligned}
    \langle \dot{p}_b \rangle &= -\frac{m v^2_\perp}{2 B_0}\frac{\partial}{\partial z} B_0(1 + \lambda z)\\
                              &= -\frac{m v^2_\perp \lambda}{2},
    \label{eq:mirror-wrong}
\end{aligned}    
\end{equation}
whereas Eq.~(\ref{eq:v-dot-para-avg}) does 
not because $\nabla \hat{b}=0$: 
\begin{equation}
    \begin{aligned}
        \langle \dot{p}_b \rangle &= m \langle v^2_\kappa \hat{\kappa}\cdot(\nabla \hat{b}\cdot \hat{\kappa})\rangle 
                                  + m\langle v^2_\tau \hat{\tau}\cdot(\nabla \hat{b}\cdot \hat{\tau})\rangle \\
                                  &= 0.
    \label{eq:mirror-right}
\end{aligned}    
\end{equation}
Indeed, there cannot be a mirror effect because the $z$-component of the
Lorentz force is always zero.

\section{Gradient-B drift}
\label{sec:grad-b-drift}

STUDENT: The gradient-B drift should be the orbit average of the
field-magnitude driver $\mathbf{v}_{\nabla B}(\mathbf{v})$ from
Eq.~(\ref{eq:field-magnitude-velocity-driver}).

INSTRUCTOR: Try it.

STUDENT: There are two pieces. Take the Lorentz cross-product
    piece $(m/qB_0^2)\mathbf{v}\times\hat{b}(\mathbf{v}\cdot\nabla B)$ first.
    In Frenet-Serret coordinates the field gradient is
\begin{equation}
    \nabla B = \frac{\partial B}{\partial \kappa} \hat{\kappa} + \frac{\partial B}{\partial \tau}\hat{\tau} + \frac{\partial B}{\partial b} \hat{b},
    \label{eq:grad-b-fs}
\end{equation}
the velocity is $\mathbf{v} = v_\kappa\hat{\kappa} + v_\tau\hat{\tau} +
v_b\hat{b}$ (Eq.~\ref{eq:particle-velocity}), and $\mathbf{v}\times\hat{b} =
v_\tau\hat{\kappa} - v_\kappa\hat{\tau}$. Expanding,
\begin{equation}
\begin{aligned}
    \frac{m\mathbf{v}\times\hat{b}(\mathbf{v}\cdot\nabla B)}{qB_0^2} &= \frac{m}{qB_0^2}\left(v_\kappa v_\tau \frac{\partial B}{\partial \kappa} + v_\tau^2 \frac{\partial B}{\partial \tau} + v_\tau v_b \frac{\partial B}{\partial b}\right)\hat{\kappa} \\
    &- \frac{m}{qB_0^2}\left(v_\kappa^2 \frac{\partial B}{\partial \kappa} + v_\kappa v_\tau \frac{\partial B}{\partial \tau} + v_\kappa v_b \frac{\partial B}{\partial b}\right)\hat{\tau}.
    \label{eq:field-magnitude-velocity-specific}
\end{aligned}
\end{equation}

The zeroth-order motion is circular gyration, so $\langle
    v_\kappa\rangle = \langle v_\tau\rangle = 0$, $\langle v_\kappa
    v_\tau\rangle = 0$, and $\langle v_\kappa^2\rangle = \langle
    v_\tau^2\rangle = v_\perp^2/2$. Only the $v_\kappa^2$ and $v_\tau^2$ terms
    survive, with the gradient pulled out of the average (first order in
    $\varepsilon$):
\begin{equation}
\begin{aligned}
    \left\langle\frac{m\mathbf{v}\times\hat{b}(\mathbf{v}\cdot\nabla B)}{qB_0^2}\right\rangle &= -\frac{mv_\perp^2}{2qB_0^2}\left(\frac{\partial B}{\partial \kappa}\hat{\tau} - \frac{\partial B}{\partial \tau}\hat{\kappa}\right) \\
    &= -\frac{mv_\perp^2}{2qB_0^2}\hat{b}\times\nabla B.
\end{aligned}
\label{eq:grad-b-first-piece}
\end{equation}
That is exactly minus the gradient-B drift.

INSTRUCTOR: The other piece?

STUDENT: That's the $\delta B$ piece $-(2\delta B/B_0 + \delta
    B^2/B_0^2)\mathbf{v}_\perp$. I need an expression for $\delta B$. It is the
    field magnitude at the particle's position minus the value at the
    gyrocenter, so at first order in $\varepsilon$,
\begin{equation}
    \delta B \approx \boldsymbol{\rho}\cdot\nabla B,
\end{equation}
where $\boldsymbol{\rho} \equiv \mathbf{r} - \mathbf{R}$ is the displacement
from the gyrocenter. At zeroth order the particle moves in a circle of radius
$r_L$ with $\boldsymbol{\rho}$ perpendicular to $\mathbf{v}_\perp$:
\begin{equation}
    \boldsymbol{\rho} = -\frac{m}{qB_0}\mathbf{v}\times\hat{b},
    \label{eq:gyroradius-vector}
\end{equation}
which has magnitude $r_L$ and can be verified by differentiating to recover
$\dot{\boldsymbol{\rho}} = \mathbf{v}_\perp$. Substituting,
\begin{equation}
    \delta B = -\frac{m}{qB_0}(\mathbf{v}\times\hat{b})\cdot\nabla B.
    \label{eq:deltaB-first-order}
\end{equation}

The size is $|\delta B/B_0| \leq \varepsilon$, the oscillation is
    at the gyrofrequency, and $\langle\delta B\rangle = 0$. Dropping $\delta
    B^2/B_0^2$ as second-order, the $\delta B$ piece becomes
\begin{equation}
    -\frac{2\delta B}{B_0}\mathbf{v}_\perp = \frac{2m}{qB_0^2}\left[(\mathbf{v}\times\hat{b})\cdot\nabla B\right]\mathbf{v}_\perp.
    \label{eq:deltaB-piece}
\end{equation}
In Frenet-Serret coordinates $(\mathbf{v}\times\hat{b})\cdot\nabla B =
v_\tau(\partial B/\partial\kappa) - v_\kappa(\partial B/\partial\tau)$, and the
same gyrotropic averages collapse the product to
\begin{equation}
    \left\langle\left[(\mathbf{v}\times\hat{b})\cdot\nabla B\right]\mathbf{v}_\perp\right\rangle = \frac{v_\perp^2}{2}\hat{b}\times\nabla B,
\end{equation}
giving
\begin{equation}
    \left\langle-\frac{2\delta B}{B_0}\mathbf{v}_\perp\right\rangle = +\frac{mv_\perp^2}{qB_0^2}\hat{b}\times\nabla B,
    \label{eq:grad-b-second-piece}
\end{equation}
exactly $-2$ times the Lorentz cross-product piece. The two pieces sum to
\begin{equation}
    \langle\mathbf{v}_{\nabla B}(\mathbf{v})\rangle = \frac{mv_\perp^2}{2qB_0^2}\hat{b}\times\nabla B,
    \label{eq:grad-b-drift-final}
\end{equation}
the standard gradient-B drift, with the correct sign.

INSTRUCTOR: What's the $\delta B$ piece doing physically?

STUDENT: The Larmor radius depends on the local field magnitude,
which the particle samples as it gyrates. The rate of change of that
magnitude is captured by the convective derivative in
Eq.~(\ref{eq:field-magnitude-velocity-driver}). By introducing $\delta B$
we shift the reference point from the particle to the center of gyration, which
allows us to treat the dynamics as a driven oscillator: clean gyration
at $\Omega_0$, perturbed by variations in $B$ and $\hat{b}$.

\section{Concluding remarks} \label{sec:conclusion}

The first-order motions derived here are mathematically equivalent to those
derived in standard treatments\cite{Kruskal1965}. Separating the convective
derivative in Eq.~(\ref{eq:b-dot}) into field-direction and field-magnitude
parts (Eq.~\ref{eq:B-grad-split}) provides a natural framework for explaining the three
canonical motions of charged particles in static non-uniform magnetic fields.
The curvature drift and mirror effects come from the rotation of the field
along the particle trajectory. Identifying the specific source terms in the
equations of motion provides physical insight that the gyroaveraged
explanations cannot. This includes the mirror force as an effect of the
gradient in the field direction rather than magnitude, and the case of
field-aligned motion for which the concepts of gyromotion and the invariance of
the magnetic moment don't apply. 

INSTRUCTOR: What causes the magnetic curvature drift?

STUDENT: It's the Lorentz force. Imagine a particle that is moving exactly
parallel to the curving field. Inertia carries it into a region where the magnetic field direction is
different. The Lorentz force rotates the velocity vector. The greater the
curvature, the more the angle between the velocity and the field widens, making
the Lorentz force stronger, which pushes the particle back into alignment with
the curving field. The velocity rotation isn't symmetric. That's the curvature
drift. It happens \textit{because} the particle isn't able to follow the
magnetic field line!

\begin{acknowledgments} 

The author thanks Abdelhaq Hamza for pointing out the Kruskal tutorial.
Critical feedback from two reviewers led to a significantly improved
manuscript.  The author appreciates fun and insightful discussions about
charged-particle motion with Eric Donovan, Larson Scullion, and students in the
University of Calgary's Winter 2026 PHYS 509 course Plasma Physics.

\end{acknowledgments}

\end{document}